\documentclass[prstab,reprint,showpacs,tightenlines, showkeys,superscriptaddress,amssymb,twocolumn]{revtex4-1}
\usepackage{amsmath,upgreek,graphicx,tikz}
\usepackage{subfigure}
\usepackage{color}
\usepackage{soul}
\usepackage{float}
\usepackage{lipsum}
\usepackage{comment}
\begin{document}

\title{Nonparametric Modeling of Face-Centered Cubic Metal Photocathodes}
\date{\today}

\author{Tuo Li}
\email{lizanchen1986@gmail.com}
\affiliation{Uptake Technologies, Inc., 600 W Chicago Ave Ste 620, Chicago, IL 60654, USA}
\author{W. Andreas Schroeder}
\affiliation{Department of Physics, University of Illinois at Chicago, 845 W. Taylor Street, Chicago,
IL 60607-7059, USA}

\begin{abstract} 
Face-centered cubic (FCC) is an important crystal structure, and there are ten elemental FCC metals (Al, Ag, Au, Ca, Cu, Pb, Pd, Pt, Rh, and Sr) that have this structure. Three of them could be used as photocathodes (Au, Rh, Pt, and Pd have very high work functions; Ca and Sr are very reactive). Au has high work function, but it is included for the sake of the completeness of noble metals' photoemission investigation. In this paper, we will apply the nonparametric photoemission model to investigate these four remaining FCC photocathodes; two noble metals (Cu and Au), two $p$-block metals (Al and Pb). Apart from the fact that the direct photoemission is dominant for most FCC photocathodes, photoemission from a surface state has also been observed for the (111)-face of noble metals. The optical properties of the (111) surface state will be extensively reviewed both experimentally and theoretically, and a surface state DFT evaluation will be performed to show that the photocathode generated hollow cone illumination (HCI) can be realized.\\
\end{abstract}


\maketitle

\section{Introduction} 
Noble metal photocathodes with low emittance (high brightness) are nowadays routinely used as electron sources for laser-driven high-gradient guns~\cite{berger_excited-state_2012, schmerge_rf_2007, nicolay_spin-orbit_2001, paniago_temperature_1995, el-fattah_modifying_2012, lashell_spin_1996, hsieh_resonances_1987, burgi_noble_2000, smith_phase_1985}. That is because these metals are resistant to corrosion, easy to fabricate, reliable and tolerant to contamination, and have uniform emission, high operating life time and a fast response time. The other principal reason for the wide use of noble metals as photocathodes is, of course, the wealth of available information about them. For example, the electronic properties of noble metals have been thoroughly investigated by cyclotron resonance, XPS experiments, X-ray emission experiments, ARUPS experiments and de Haas-van Alphen measurements~\cite{koch_new_1964, courths1984photoemission, joseph_low-field_1966}. Each experimental method mentioned above has been extensively applied to the field of photoemission physics and the results have been found to fit a consistent and entirely reasonable picture of the metal's Fermi surface. The disadvantages of noble metal photocathodes are their low QE (of the order of $10^{-6}\sim10^{-4}$) and the need for a UV drive laser ($>$ 4.60eV), which limits these photocathodes for applications requiring less than $\approx$ 1mA average current~\cite{dowell_cathode_2010}.\

In recent years, axial hollow cone illumination has gained great interest at high resolutions in conventional transmission electron microscopes since it eliminates phase contrast artifacts from an image, thus making high resolution electron microscopy (HREM), bright-field imaging and dark-field electron microscopy (DFEM) much more reliable~\cite{krakow_method_1976, dinges_high-resolution_1994, kondo_new_1984}. The current method of producing HCI employs annular condenser apertures, circular condenser apertures, or an electronic cone illumination method, which impose several instrumental limitations. It is therefore important to consider an alternative method of producing hollow cone illumination which minimizes such limitations and enables the corresponding microscopy to be performed routinely. Aside from the needed instrument improvements, it is inevitable that the standard methods of producing cone illumination mask the electron beam and therefore decrease its brightness. In this sense, a photocathode that directly generates the required hollow-cone beam for HCI would be quite attractive.\

In Section~\ref{sec:noble Metal Photocathodes}, we will conduct the nonparametric DFT-based analysis for Ag, Cu and Au. In Section~\ref{sec:Hollow Cone Illumination}, we will present HCI based on Ag(111) single crystal emission, and show the schematic procedure to generate a `cone-like' electron beam. As $\Delta p_T$ data on photoemission from the Ag(111) face surface state is lacking in the photocathode community, our DFT calculation~\cite{li2015emission,w2015photocathode,w2015oriented,schroeder2014metal} fills this vacancy. Emphasis will be placed on a discussion of the surface state photoemission and the resulting $\Delta p_T$ analysis and results obtained with it. In Section~\ref{sec:summary and discussion}, the DFT-based photoemission analysis is used for the evaluation of the emission properties of surface states.\

\section{DFT-Based photoemission analysis} \label{sec:noble Metal Photocathodes}
\begin{figure}[h]
\center
\includegraphics[scale=0.8]{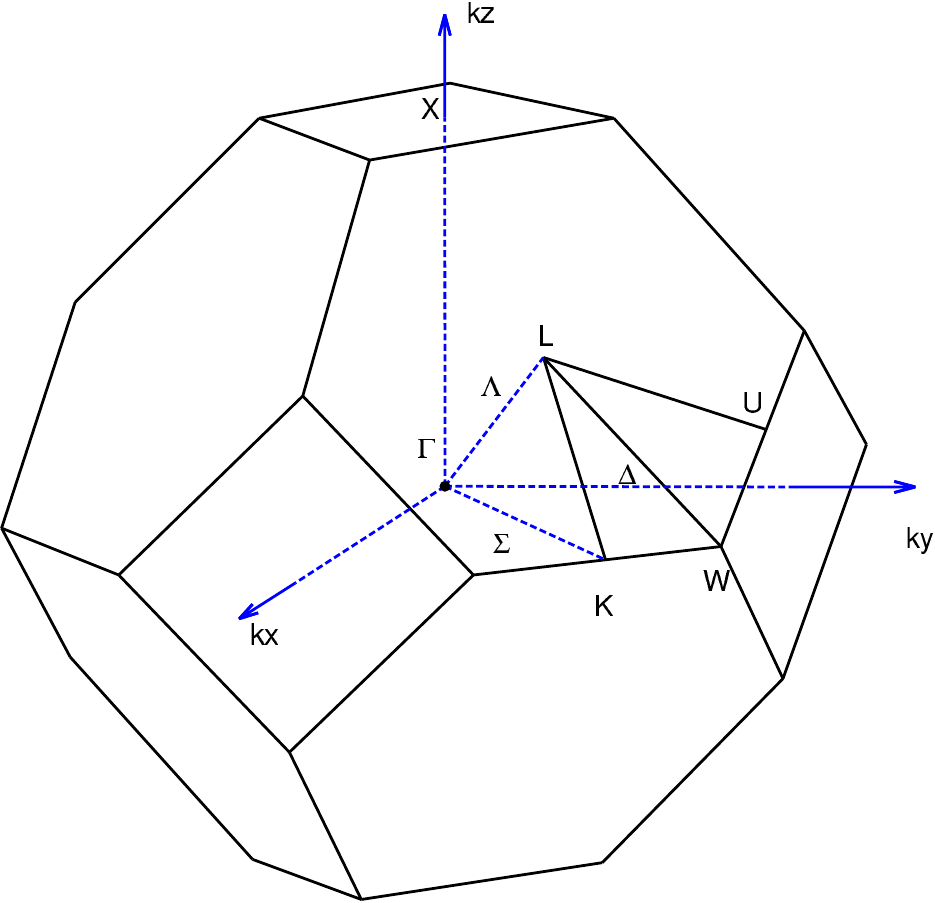}\\
\caption{FCC Brillouin zone}
\label{fig:fccbz.pdf}
\end{figure}
\begin{figure}[htb]
\hspace{0em}
\centering
  \begin{tabular}{@{}l@{}}
\includegraphics[scale=0.70]{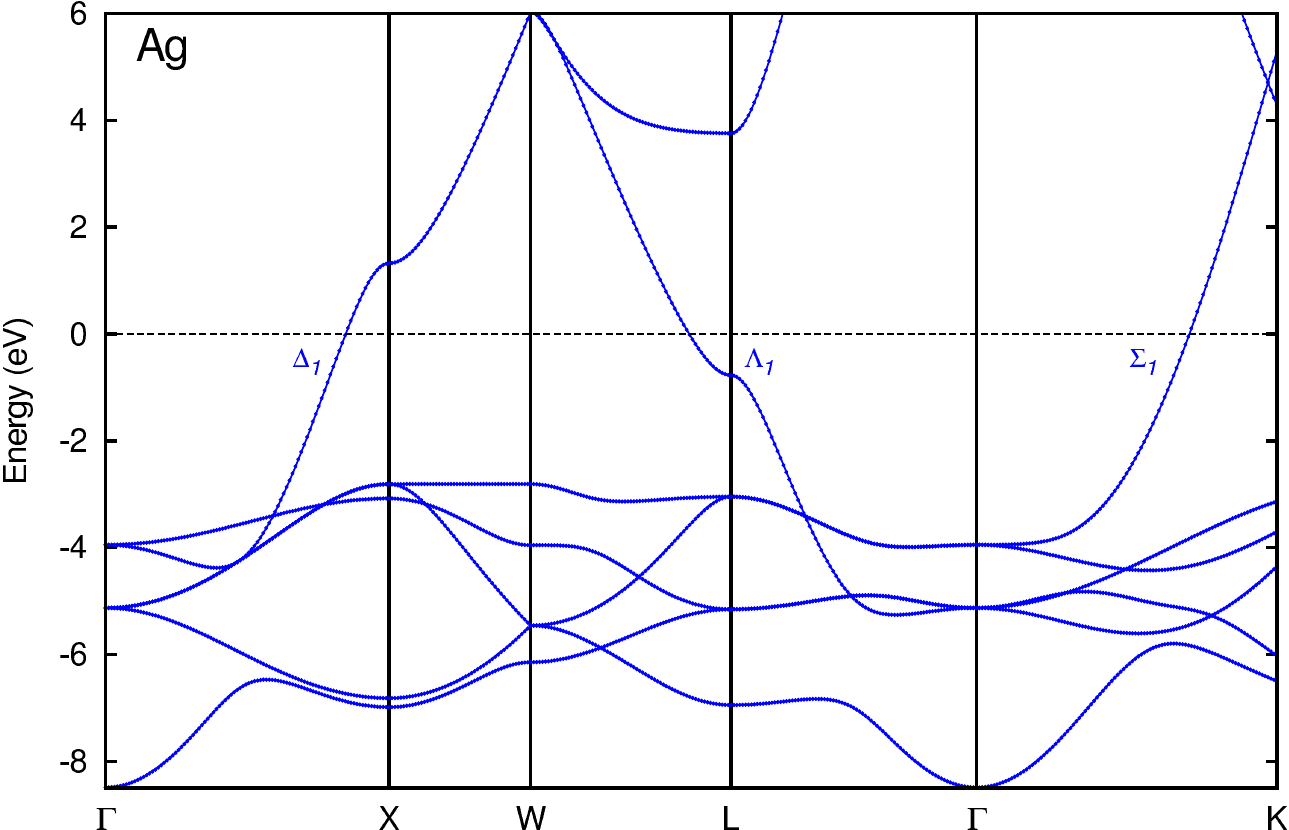}\\
\includegraphics[scale=0.70]{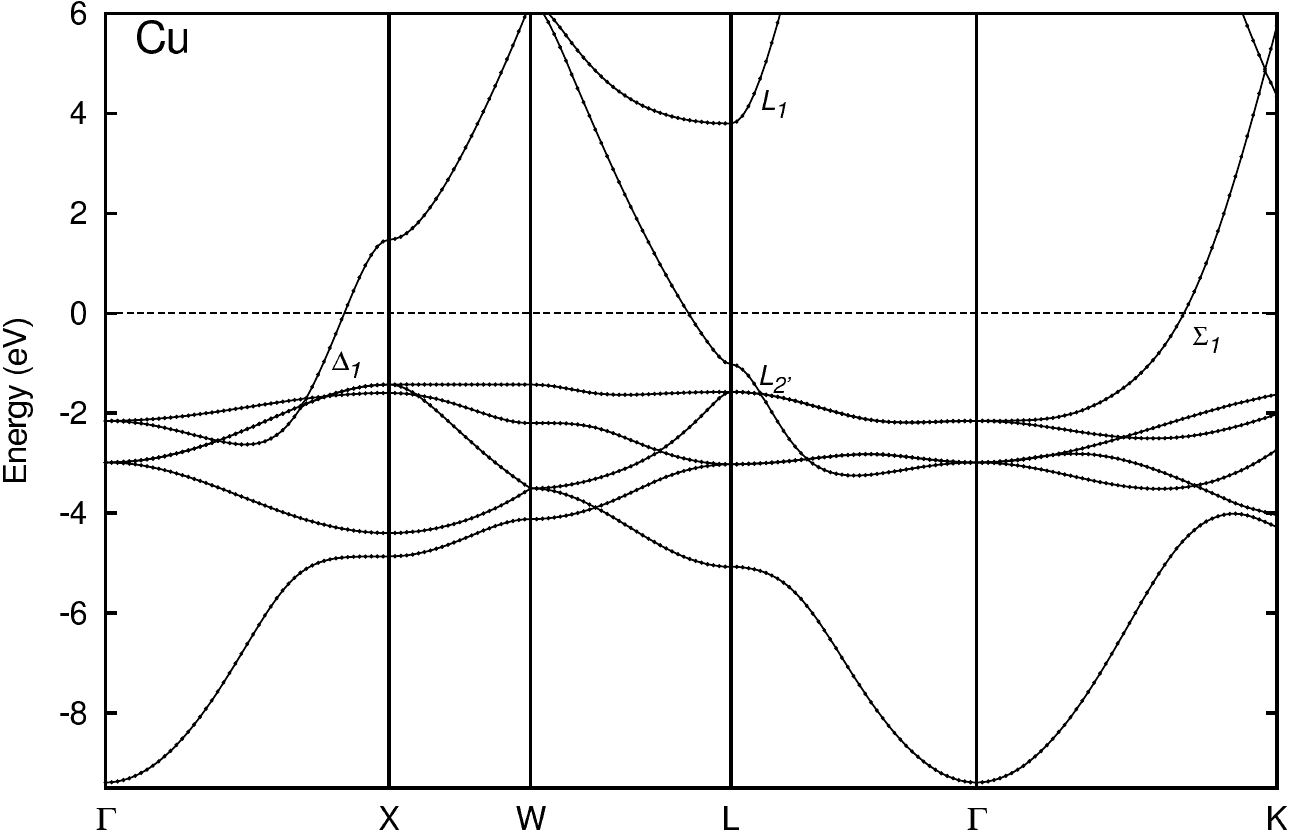}\\
\includegraphics[scale=0.70]{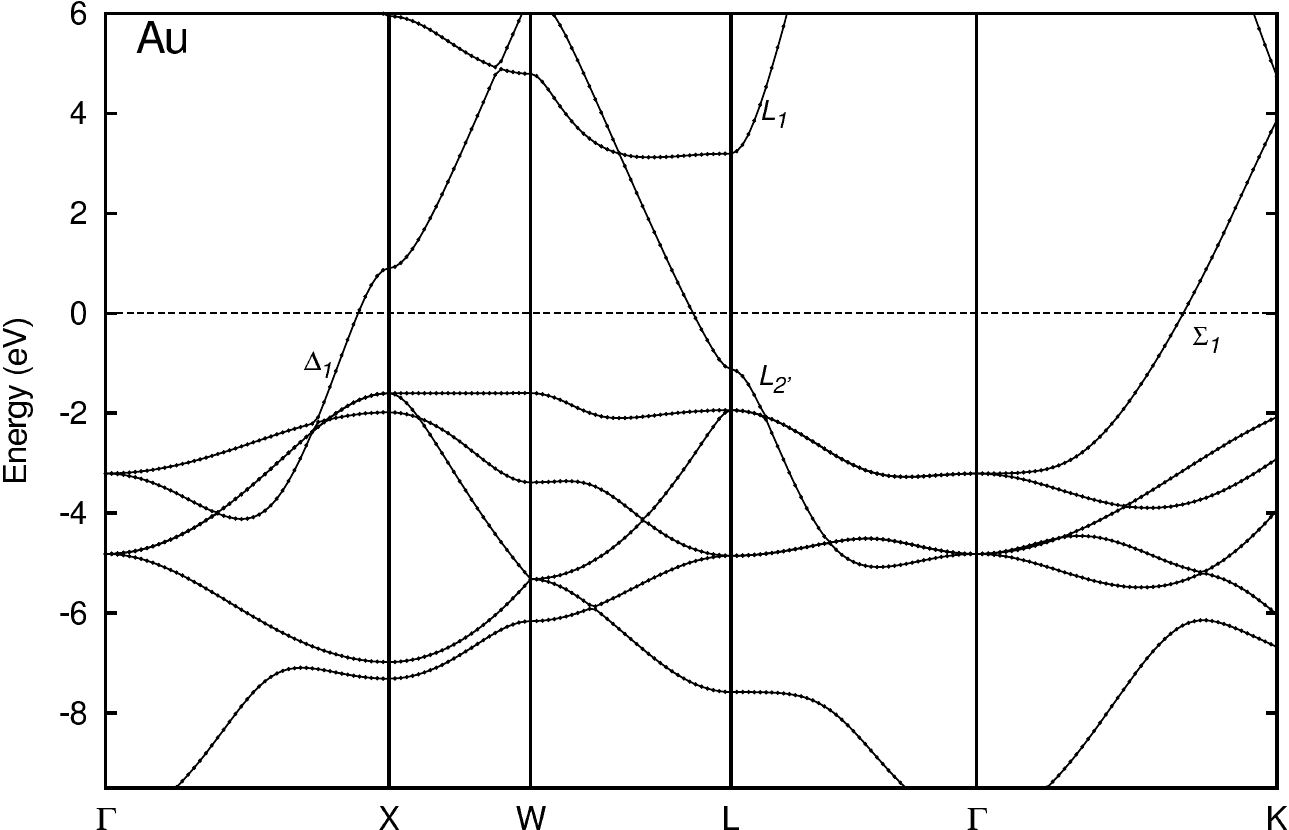}\\
  \end{tabular}
\caption{Silver, Copper and Gold band structures along high symmetry points and lines. The Fermi level is shifted to zero energy and labeled with dashed line. In [001] ($\Gamma_1\rightarrow$ X) and [110] ($\Gamma_1\rightarrow$ K) direction, $\Delta_1$ band and $\Gamma_1$ band cross Fermi level respectively. In the [111] ($\Gamma_1\rightarrow$ L) direction, $\Lambda_1$ is 0.77eV below the Fermi level.}
\label{fig:Cuband.pdf}
\end{figure}

\begin{table}[htb]
\caption{The Fermi surface radii of Ag, Al and Cu (in units of $\frac{2\pi}{a}$).}
\centering
\begin{tabular}{l l l l l} 
\hline\hline
 & face & DFT & Expt.\\ [0.2ex] 
Cu&(001)& 0.81 &0.83$^a$\\
  &(110)& 0.72 &0.74$^a$\\
Ag&(001)& 0.80 &0.82$^a$\\
  &(110)& 0.73 &0.75$^a$\\
Au&(001)& 0.87 &0.88$^a$\\
  &(110)& 0.73 &0.74$^a$\\
\hline
\end{tabular}
\label{tab:noble metal Fermi surface radii}
\begin{center}
\begin{tabular}{c}
$^a$Reference~\cite{coleridge_fermi-surface_1982}
\end{tabular}
\end{center}
\end{table}

\begin{figure}[htbp]
\centering
  \begin{tabular}{@{}ll@{}}
\includegraphics[scale=0.3]{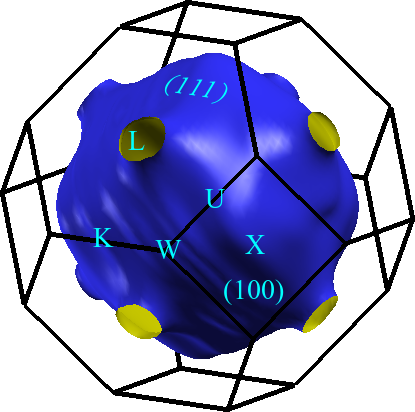}&
\includegraphics[scale=0.3]{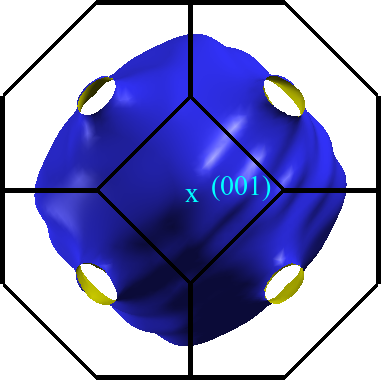}\\
\includegraphics[scale=0.3]{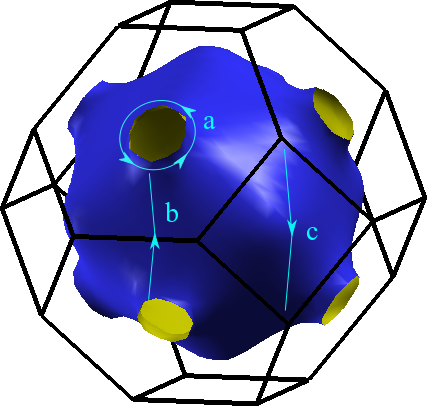}&
\includegraphics[scale=0.3]{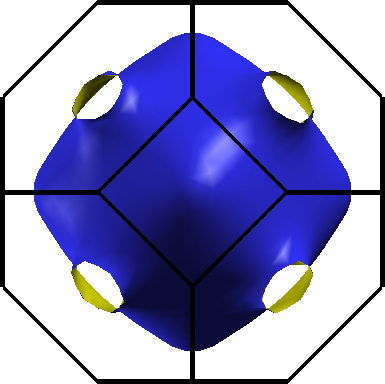}\\
\includegraphics[scale=0.3]{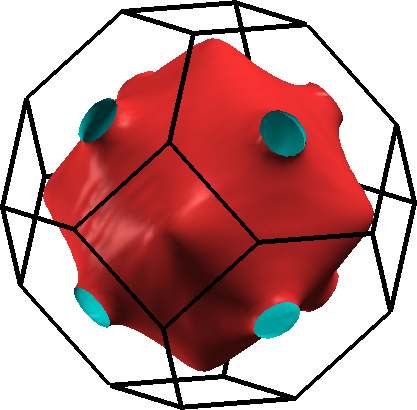}&
\includegraphics[scale=0.3]{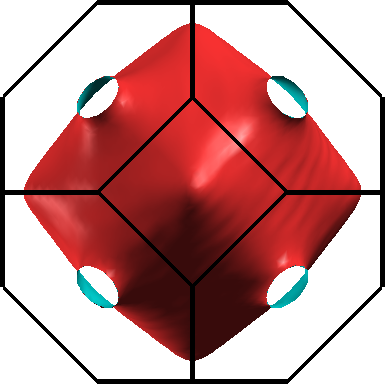}\\
  \end{tabular}
\caption{Fermi surface of Ag, Cu, and Au in first Brillouin zone. The polyhedron (black solid lines) represents the first Brillouin zone of FCC. It consists of an almost perfect `belly' sphere, centered at high symmetry point $\Gamma$, which connects to the $L$ points along the [111] direction. Two projections of the noble metals reciprocal lattice show the shape of Cu and Au Fermi surfaces. The orbits considered are labeled by light blue lines: (a) dog's bone orbit, (b) neck orbit, and (c) central belly orbit.}
\label{fig:Cufermi.png}
\end{figure}

The first Brillouin zone (BZ) of an FCC lattice has a truncated octahedron shape as shown in Fig.~\ref{fig:fccbz.pdf}. Noble metals only have one free electron per atom, so that the BZ is only half filled. The structural and electrical properties of both the bulk and surfaces of noble metals were previously investigated theoretically and experimentally~\cite{courths1984photoemission, PhysRev.136.A1030, ballinger_study_1969}. The calculated band diagrams for copper and gold are shown in Fig.~\ref{fig:Cuband.pdf}. The relatively flat bands lying about 2eV below the Fermi surface are associated principally with $d$ atomic states, whereas the bands lying at higher energy are associated principally with $s$ and $p$ atomic states. The $d$ bands are fully filled and the $s$ bands are half-filled; thus, the volume contributed by the $d$ bands is enclosed by the Fermi surface formed by the $s$ bands. The equivalent Fermi surface radius then equals the average of $\Delta_1$ ($\Gamma$--X) and $\Sigma_1$ ($\Gamma$--K) radii of the Fermi energy surface, values for which are shown in Table.~\ref{tab:noble metal Fermi surface radii} both from Ref~\cite{coleridge_fermi-surface_1982} and our DFT calculations. The unequal (100) face and (110) face radii indicate a distortion of the nominally spherical noble metal Fermi surface. In addition, contact is established between the Fermi surface and the BZ boundaries in the $<$111$>$ directions ($\overline{\Gamma L}$). These contact areas correspond to the energy gap at the zone boundary along the high symmetry line $\Lambda$. To further examine the consistency of the Fermi surface calculations by DFT, the effective masses ($m^*$) corresponding to the (001), (110) and (111) faces are determined using $m^*=(\frac{\partial^2 E(\vec{p})}{\partial^2 \vec{p}})^{-1}$. The noble metal's Fermi surface is constructed of three major parts; a `Dog's bone' orbit from the contours in the (111) plane, a `Neck' orbit from contours in the (110) plane (the hexagonal face), and a `Belly' orbit from the contours in (100) plane \cite{christensen_apw_1969, goldmann_empty_1985, joseph_low-field_1965}. These three orbits are shown in Fig.~\ref{fig:Cufermi.png} and the results from the effective masses calculations corresponding to the orbits are listed in Table.~\ref{tab:noble metal effective mass}. It can be seen that Ag, whose Fermi surface is closest to a perfect sphere, has effective mass values close to $m_0$, while Cu whose Fermi surface is furthest from a perfect sphere has an effective mass much less than $m_0$. The DFT evaluated effective masses are all within 10$\%$ of the experimental values. Unlike potassium whose band structures could be completely defined by the electron energy ($E$) versus momentum $\vec{p}$ diagram calculated along the principal crystal directions, the distorted spherical Fermi surfaces of the noble metals suggests that a DFT-based emission analysis will be required to evaluate their $\Delta p_T$.\\
\begin{table}[htbp]
\caption{The effective mass of Ag, Cu and Au (in units of $m^*/m_0$).}
\centering
\begin{tabular}{l l l l l} 
\hline\hline
 & face & DFT & Expt.\\ [0.2ex] 
\hline   
Ag& (001)&1.00&1.03$\pm$0.01$^b$\\
  & (110)&0.35&0.39$\pm$0.01$^b$\\
  & (111)&0.89&0.94$\pm$0.01$^b$\\
Cu& (100)&0.47&0.49$\pm$0.02$^a$\\
  & (110)&0.50&0.53$\pm$0.02$^a$\\
  & (111)&0.42&0.45$\pm$0.02$^a$\\
Au& (001)&0.96&0.99$\pm$0.01$^c$\\
  & (110)&0.45&0.48$\pm$0.01$^d$\\
  & (111)&1.03&1.08$\pm$0.01$^c$\\
[1ex] 
\hline 
\end{tabular}
\label{tab:noble metal effective mass}
\begin{center}
\begin{tabular}{l l}
$^a$Reference~\cite{joseph_low-field_1966}\\
$^b$Reference~\cite{lewis_band_1968}\\
$^c$Reference~\cite{joseph_low-field_1965}\\
$^d$Reference~\cite{PhysRevB.4.1197}\\
\end{tabular}
\end{center}
\end{table}
\begin{table}[htb]
\caption{The work function of Ag, Cu, and Au (in units of eV).}
\centering
\begin{tabular}{l l l l} 
\hline\hline 
 & face & Expt. & DFT\\ [0.2ex] 
\hline   
Ag&(001)&4.64$^c$&4.64$\pm$0.05\\
  &(110)&4.52$^c$&4.53$\pm$0.05\\
  &(111)&4.75$^d$&4.83$\pm$0.05\\
Cu&(100)&4.59$^b$&4.64$\pm$0.05\\
  &(110)&4.48$^b$&4.52$\pm$0.05\\
  &(111)&4.94$^b$&5.02$\pm$0.05\\
Au&(001)&5.47$^c$&5.52$\pm$0.05\\
  &(110)&5.37$^c$&5.36$\pm$0.05\\
  &(111)&5.31$^c$&5.50$\pm$0.05\\
[1ex] 
\hline 
\end{tabular}
\begin{center}
\label{tab:noble elemental metal work function}
\begin{tabular}{l l}
$^a$Reference~\cite{ma_ab-initio}\\
$^b$Reference~\cite{fall_ab_1999}\\
$^c$Reference~\cite{michaelson_work_1977}\\
$^d$Reference~\cite{PhysRev.58.812}\\
\end{tabular}
\end{center}
\end{table}
\begin{figure}[htb]
\centering
\includegraphics[scale=1.0]{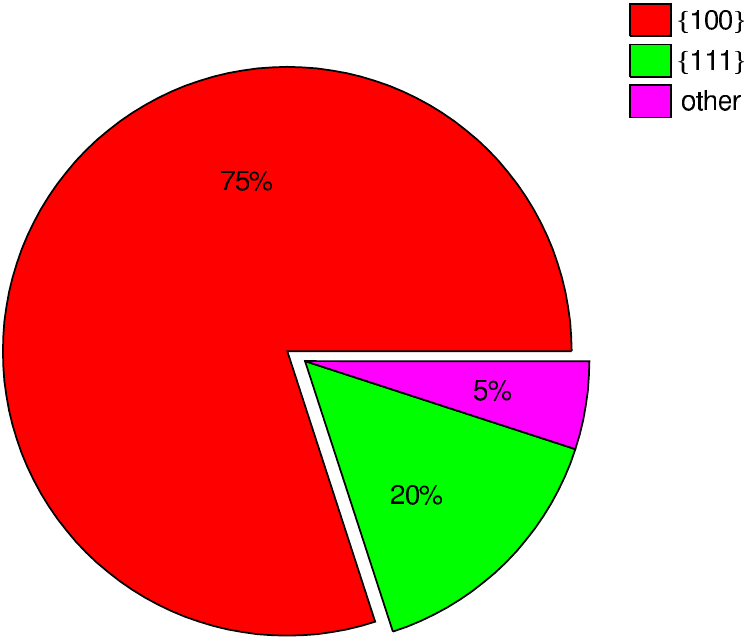}\\
\caption{General crystal orientation percentage for polycrystalline FCC metal~\cite{barrett1966structure}.}
\label{fig:bccphotomicrograph.pdf}
\end{figure}  

The results of evaluating the photoelectric work functions of noble metals with respect to crystal orientation using the thin-slab method~\cite{fall_deriving_1999}, are shown in Table.~\ref{tab:noble elemental metal work function}. The calculated work functions are within 10$\%$ of experimental values~\cite{barrett1966structure, :/content/aip/journal/jap/48/11/10.1063/1.323539}, with any absolute discrepancies being in the vicinity of 0.1eV. Nonetheless, for these metals the calculated theoretical values are consistent and appear to be reliable; indeed, they are in very good agreement with experimental values. For bulk Au, a $\hbar\omega$=5.36 eV (231 nm) photon energy is the minimum energy required to overcome the photoelectric work function barrier. Due to this high work function value, there are no literature values of the rms transverse momentum $\Delta p_T$ value for photoemission from Au with an excess energy greater than zero. For Cu and Ag, a $\hbar\omega$=4.75eV (261nm) UV laser can be used to photoemit from the (100) and (110) crystal faces. The $\Delta p_T$ of an electron beam emitted from polycrystalline Cu and Ag surfaces irradiated by the 4.75eV picosecond laser pulses has been measured by the solenoid scan technique, and the measured $\Delta p_T$ values for polished Ag and Cu photocathodes are 0.235$\sqrt{m_0eV}$ and 0.130$\sqrt{m_0eV}$, respectively. The polycrystalline nature of the measured noble metal photocathodes implies a crystal orientation randomness to the samples, so that the strict theoretical expression for $\Delta p_{T, \text{average}}$ must be expressed as
\begin{equation}
\Delta p_{T,\text{average}}=\underset{i}\Sigma\underset{j}\Sigma\underset{k}\Sigma x_{(ijk)}\Delta p_{T,(ijk)},
\end{equation}
where $x_{(ijk)}$ is the weight of $(ijk)$ crystal orientation and $\underset{i}\Sigma\underset{j}\Sigma\underset{k}\Sigma x_{(ijk)}$=1. The general $x_{(ijk)}$ distribution of FCC polycrystalline is displayed in~\ref{fig:bccphotomicrograph.pdf}, which indicates that (100) and (110) are the two prevalent crystal faces and, hence, are expected to dominate the photoemission from polycrystalline noble metals~\cite{barrett1966structure}. Therefore, it is possible to consider an average $\Delta p_{T}$ using an analysis that is restricted to (001) and (110) face emission.
\begin{figure*}[ht!]
\centering
\vspace{-1.0cm}
  \begin{tabular}{@{}ccc@{}}
\includegraphics[scale=0.3]{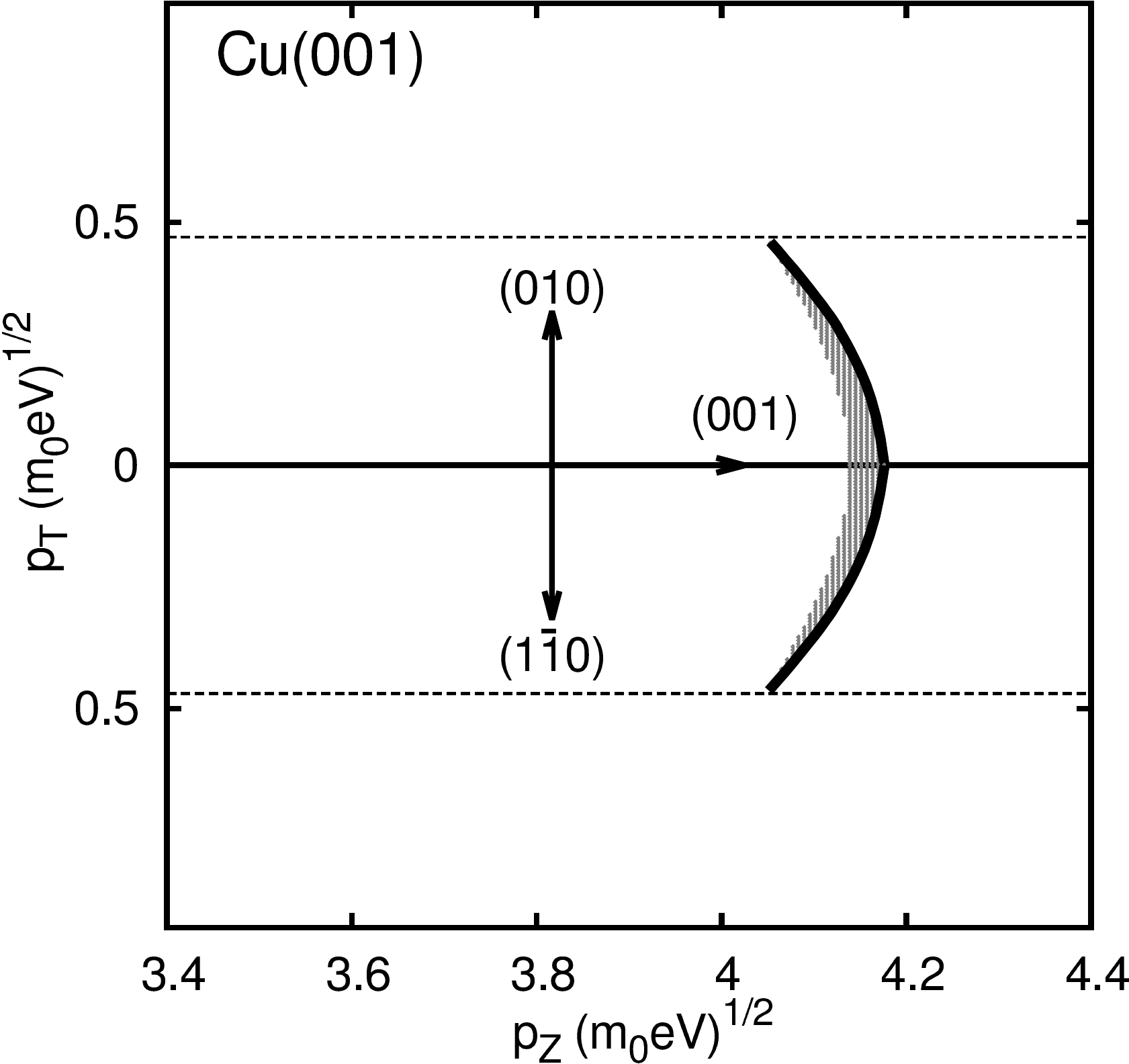}&
\includegraphics[scale=0.3]{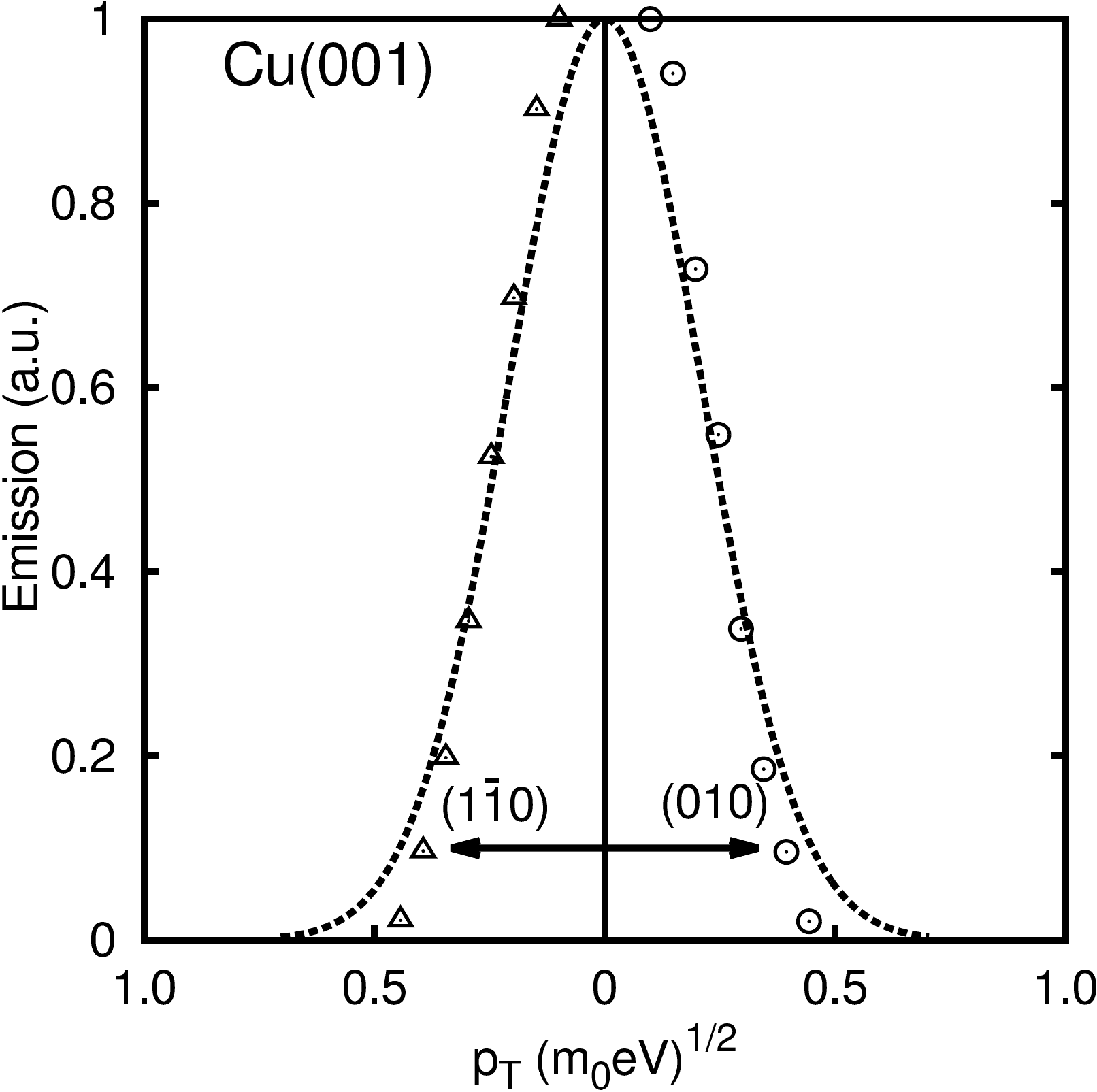}&
\includegraphics[scale=0.32]{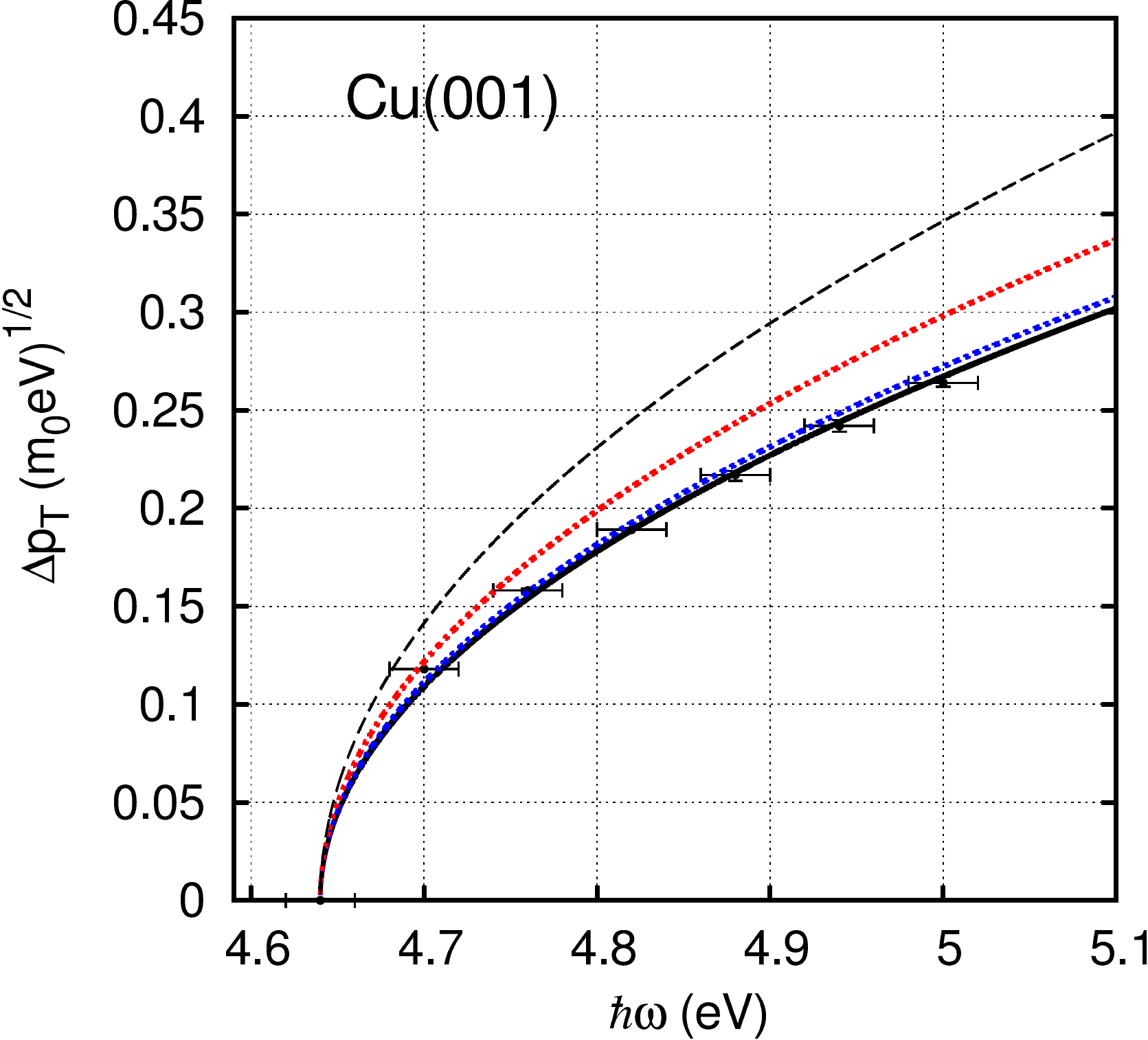}\\
\includegraphics[scale=0.3]{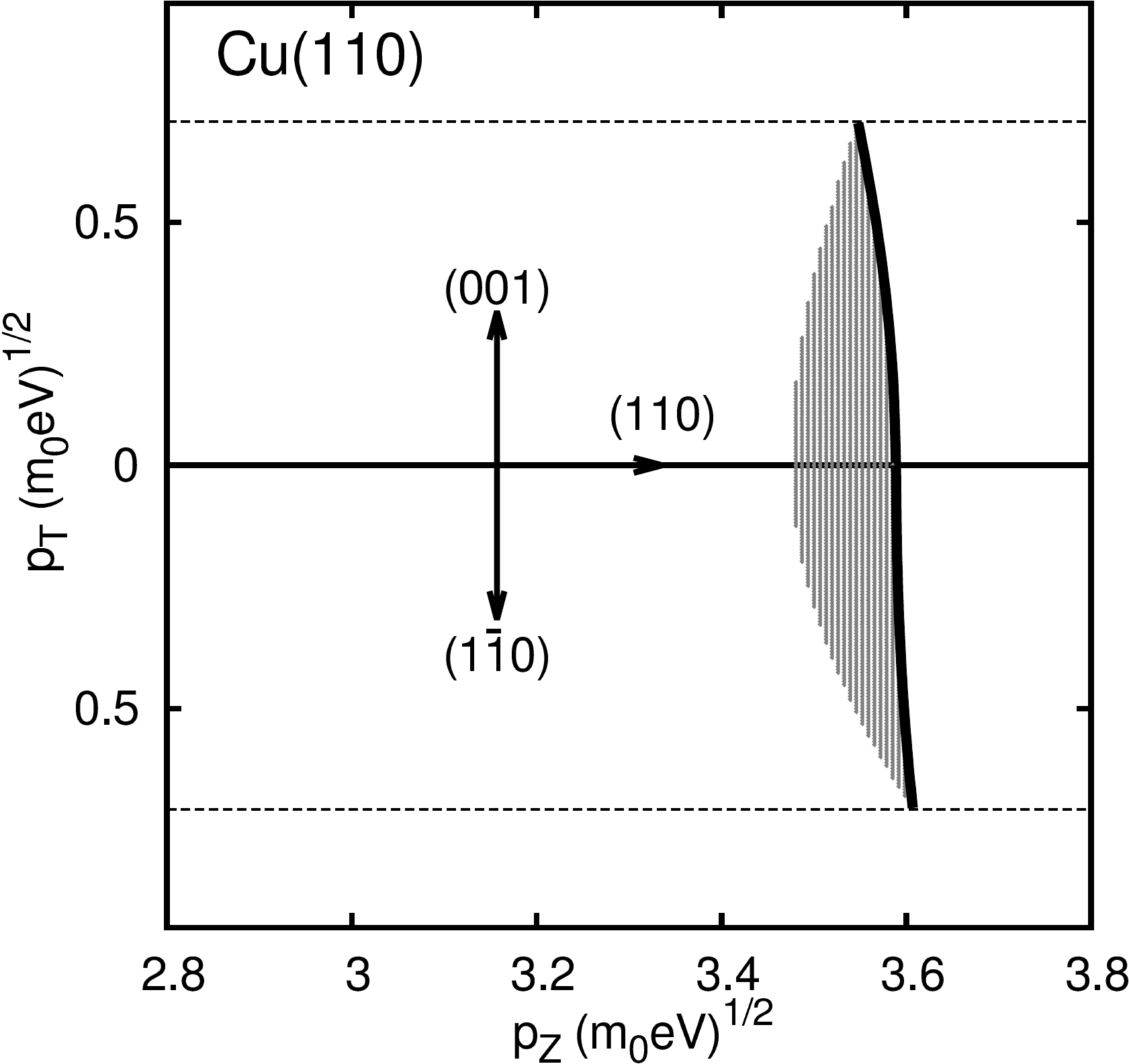}&
\includegraphics[scale=0.3]{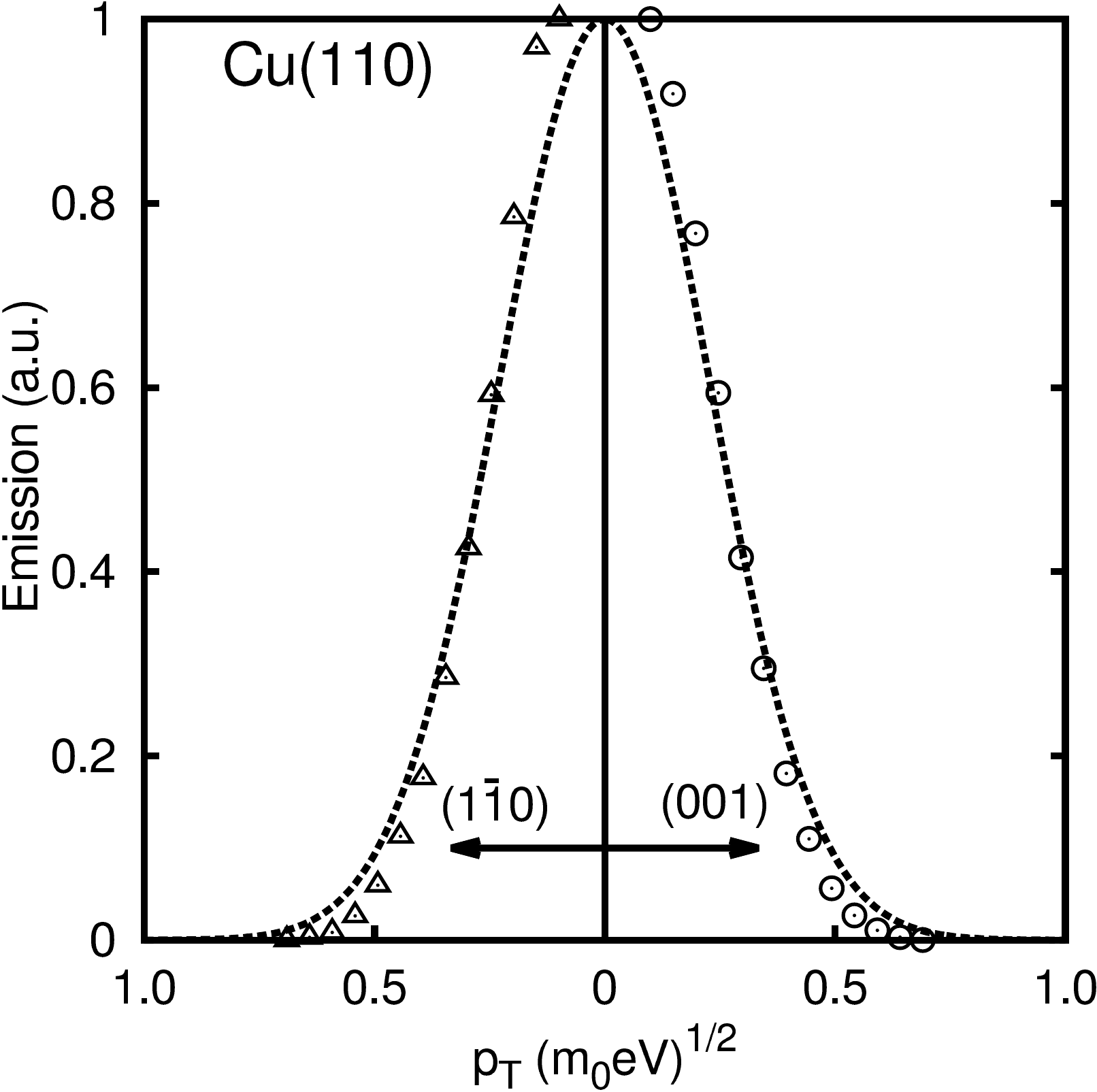}&
\includegraphics[scale=0.31]{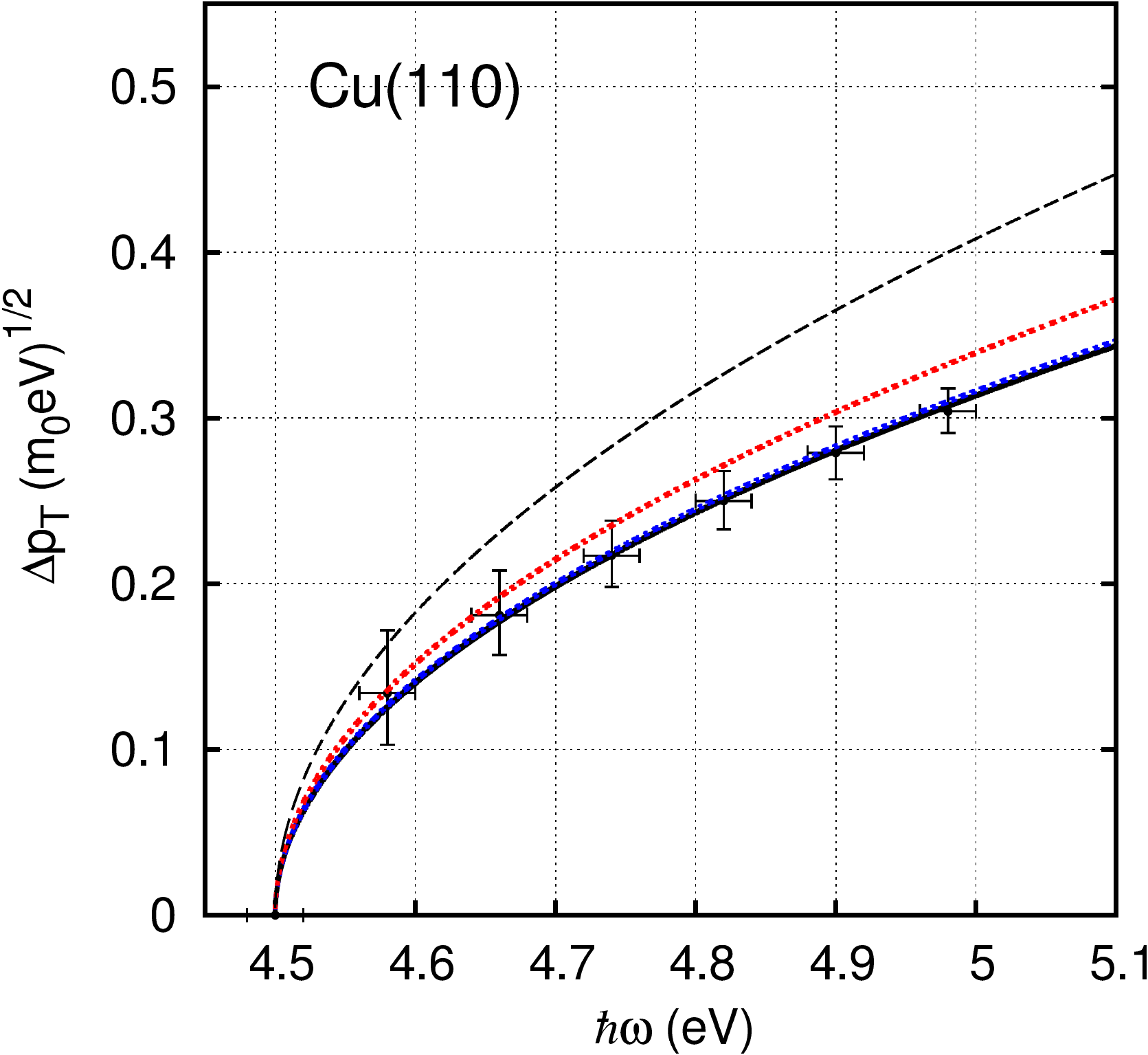}\\
\includegraphics[scale=0.3]{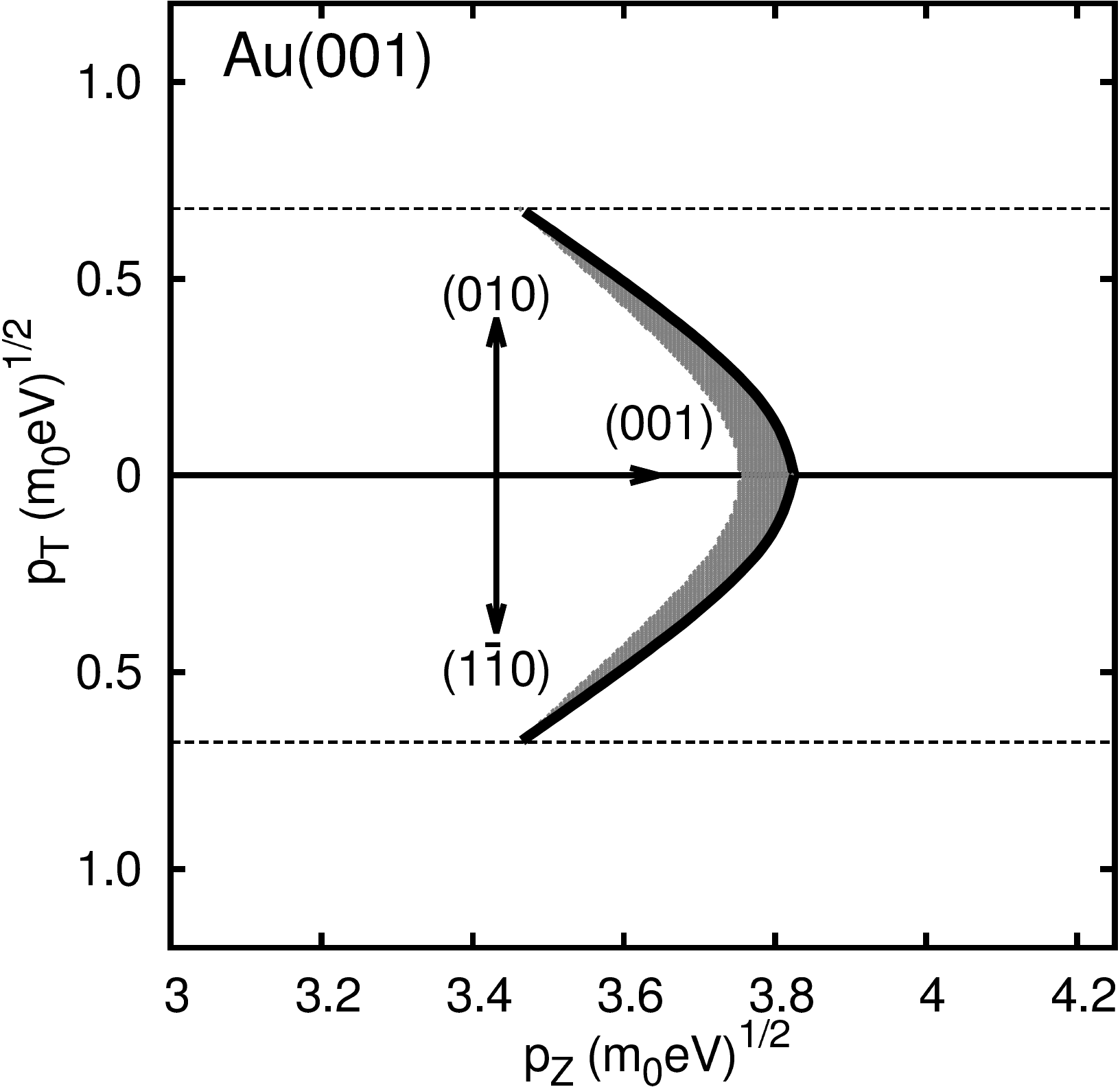}&
\includegraphics[scale=0.3]{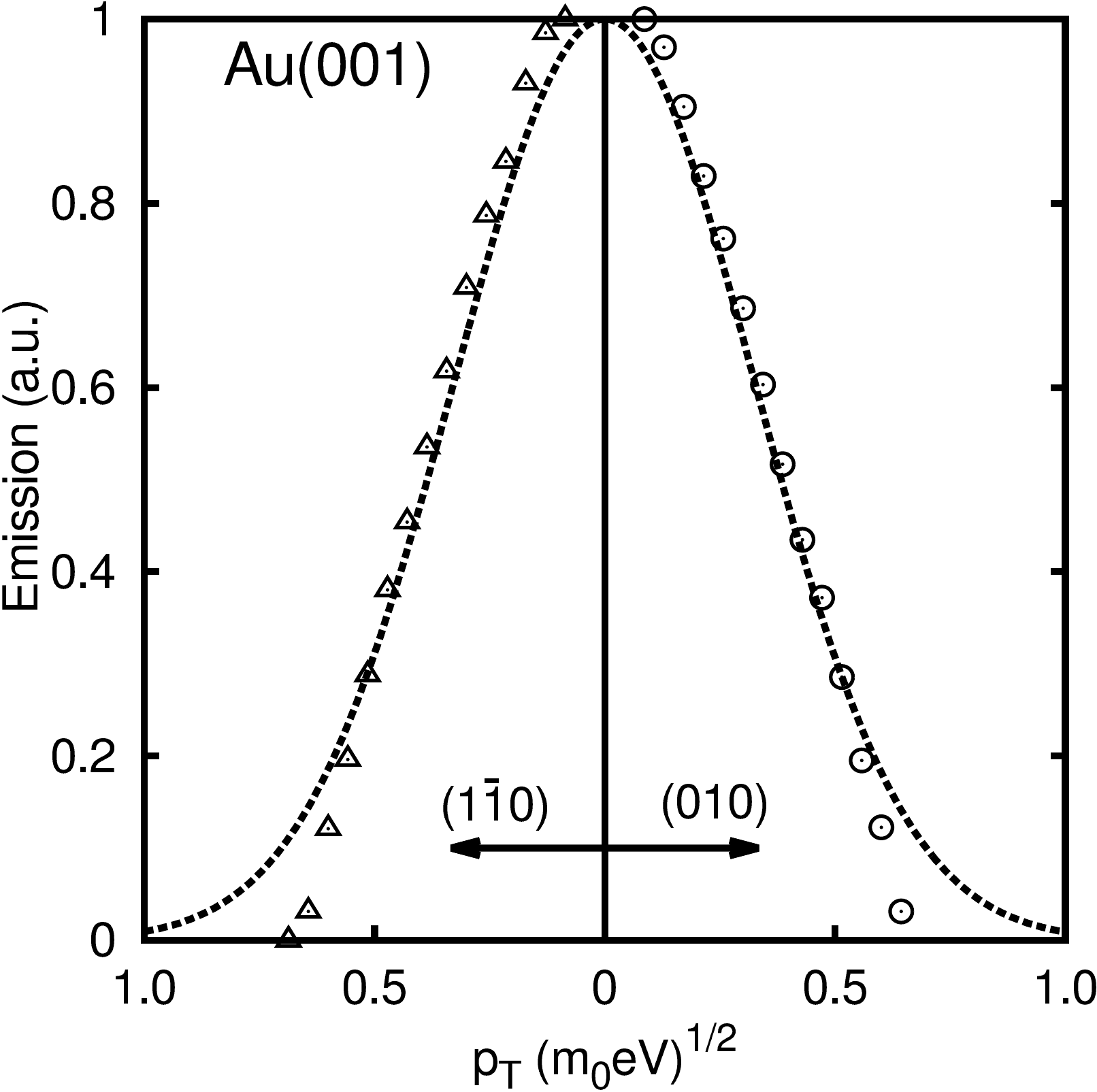}&
\includegraphics[scale=0.31]{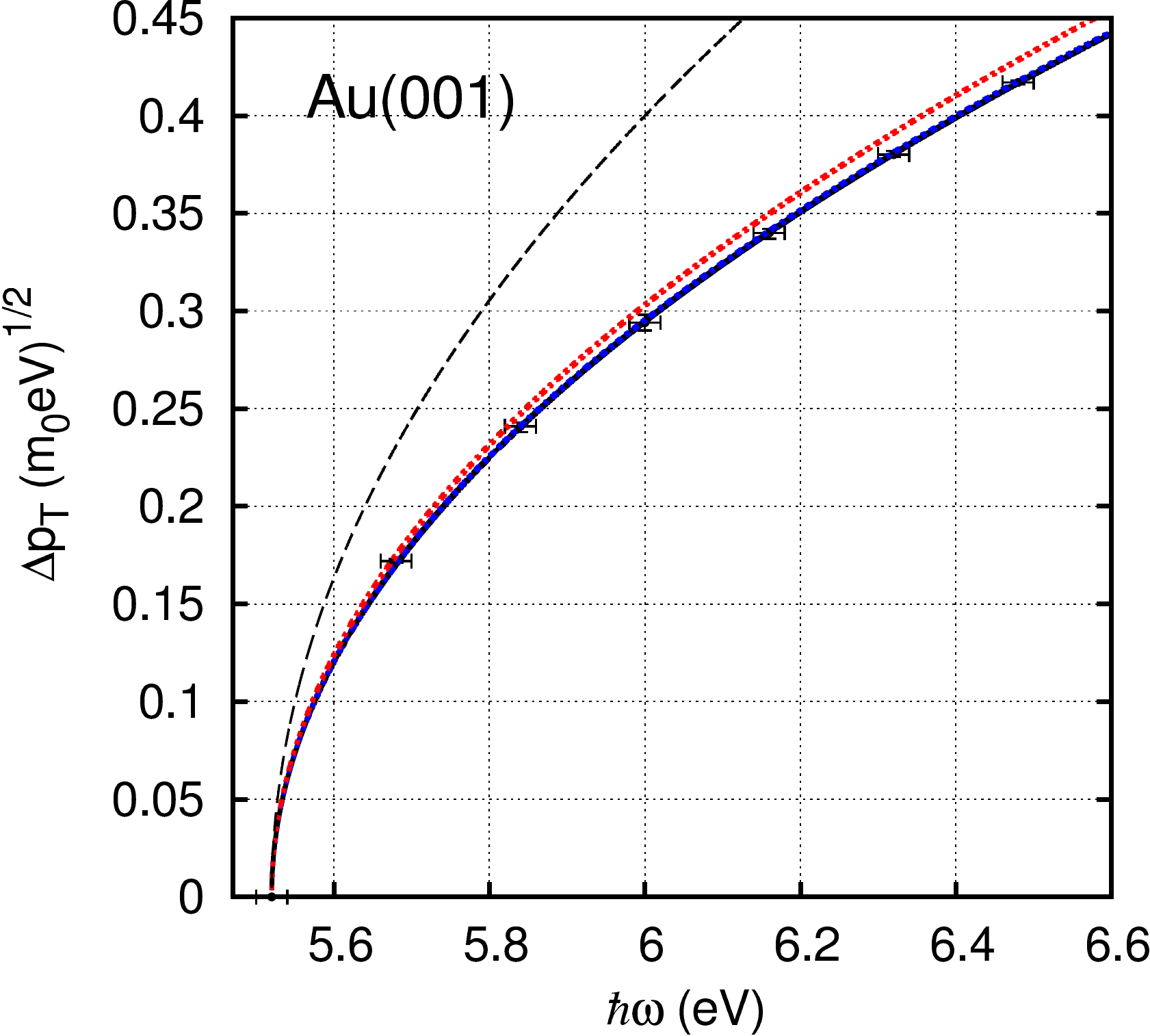}\\
\includegraphics[scale=0.3]{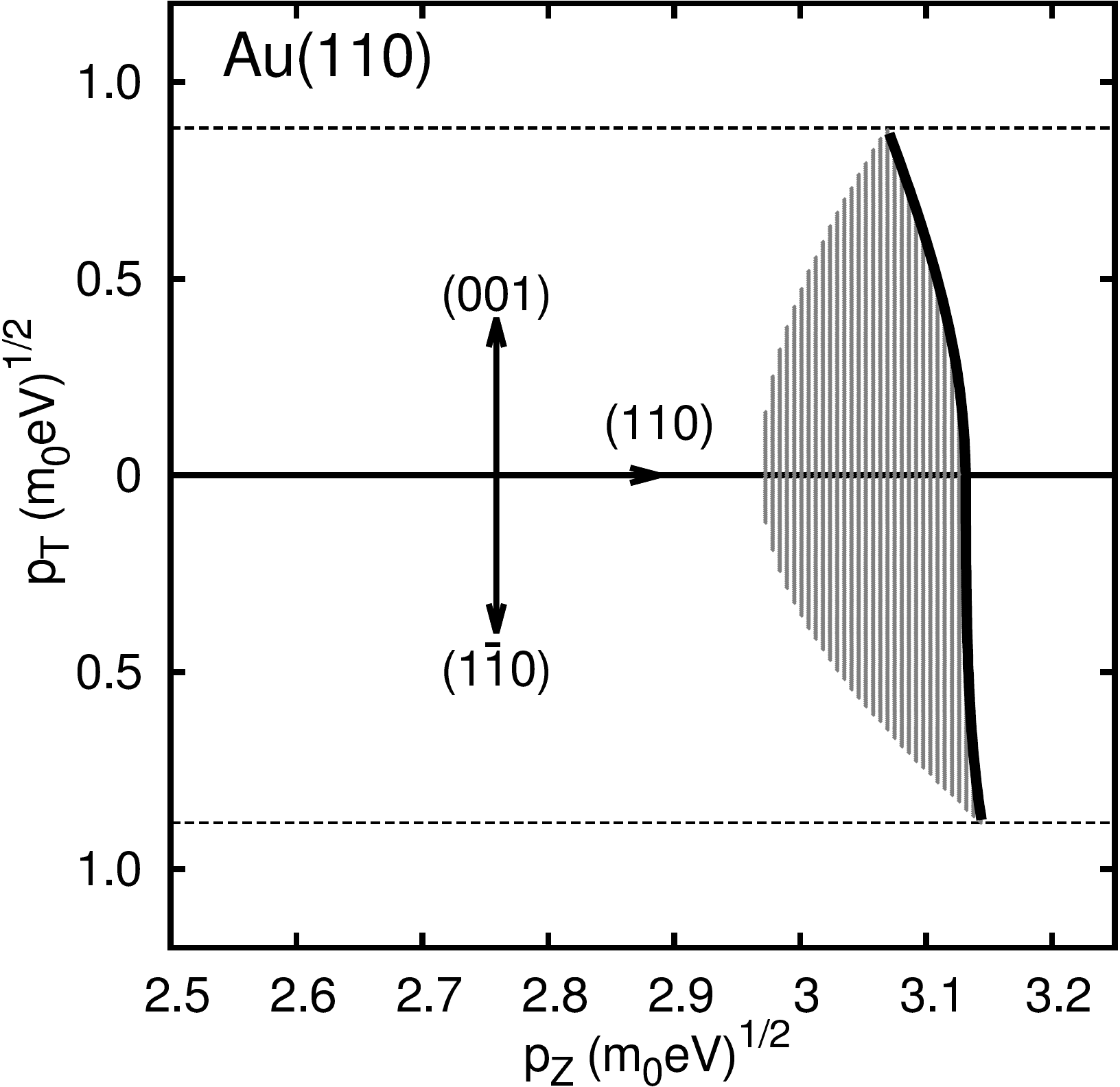}&
\includegraphics[scale=0.3]{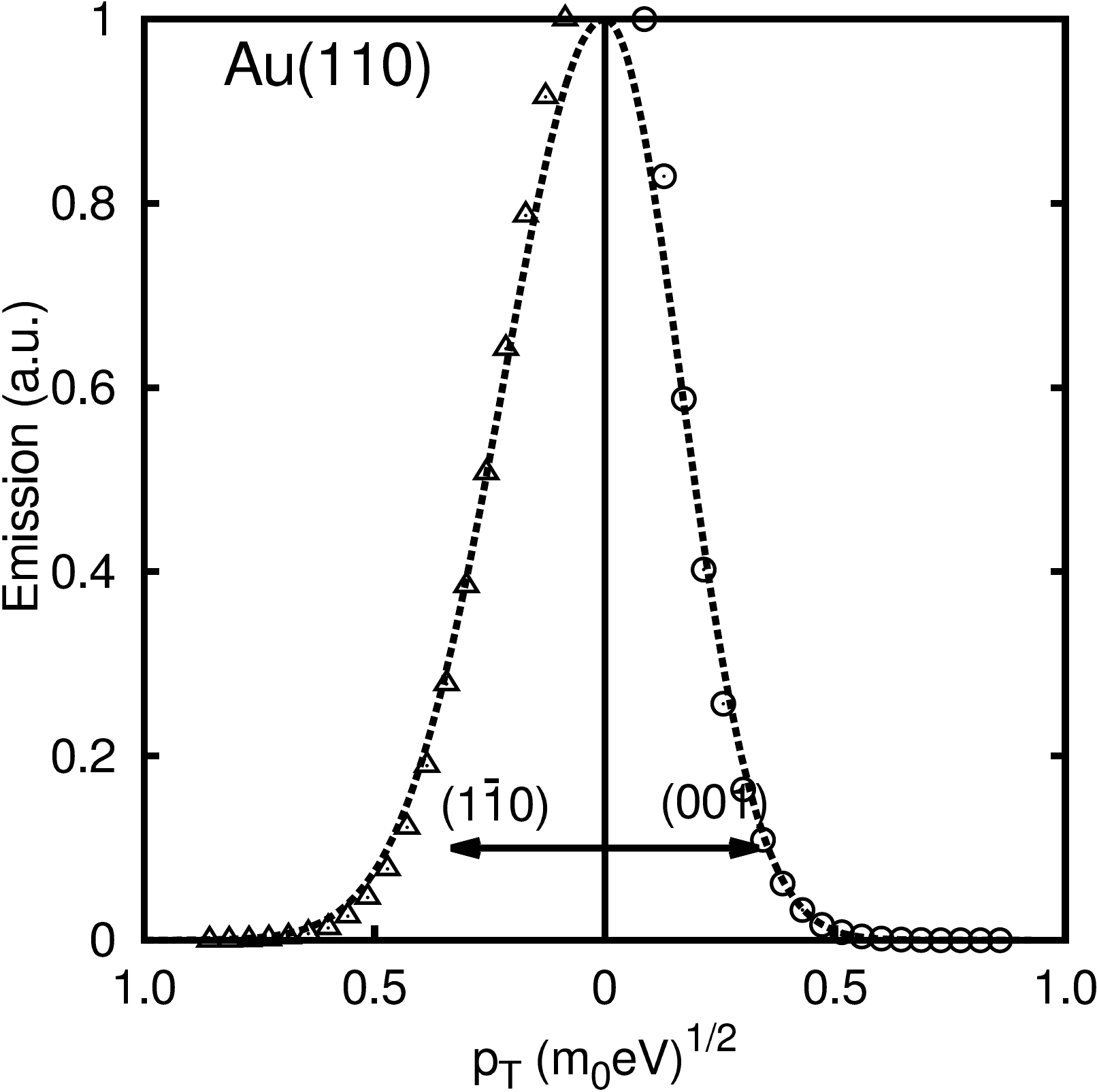}&
\includegraphics[scale=0.31]{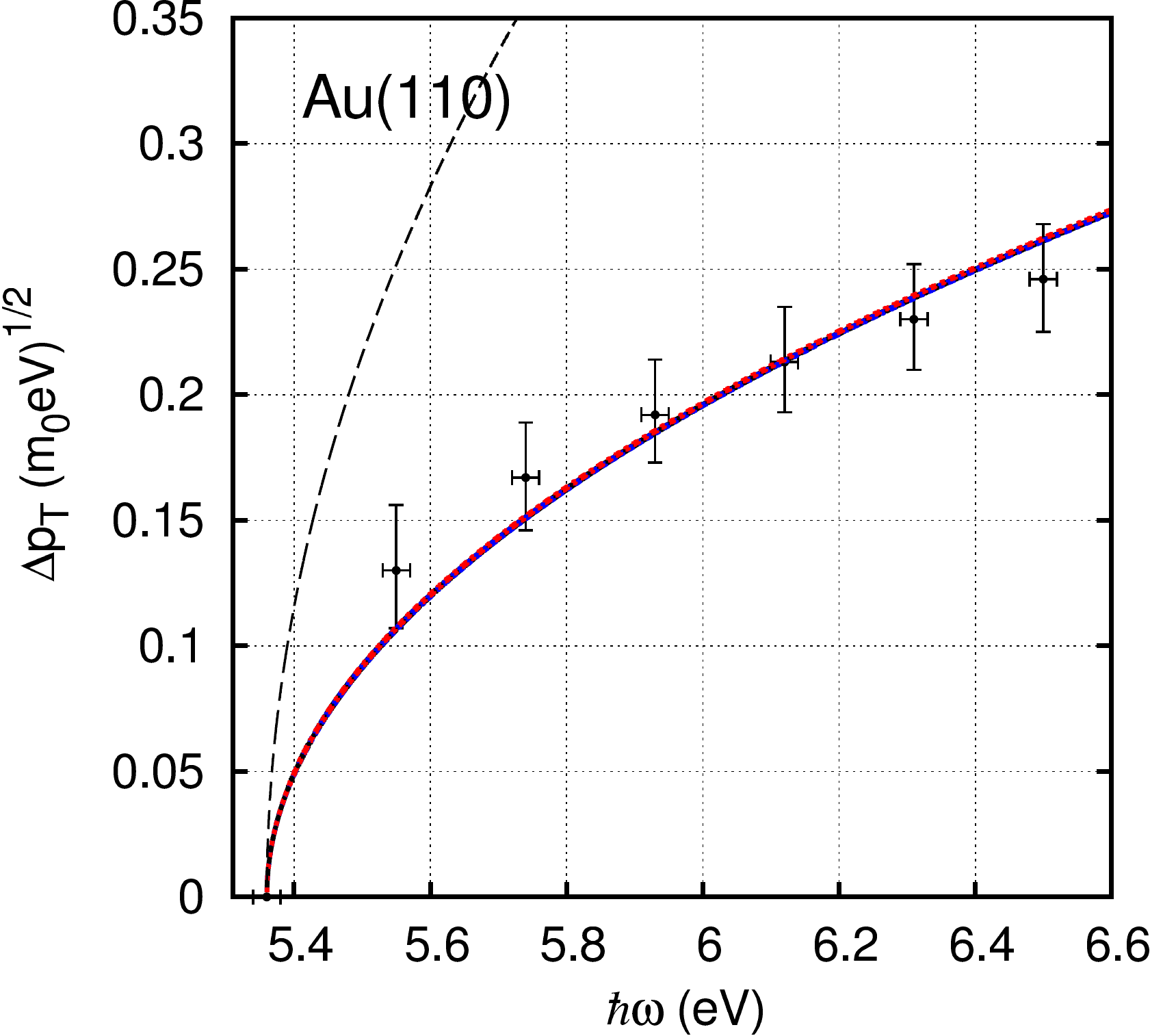}\\
(a) & (b) & (c) \\
  \end{tabular}
\caption{The results from the DFT-based photoemission analysis for the (001)/(110) face of Cu and Au. (a) Crystal momentum map of the electronic states (shaded regions) below the Fermi level (solid line) that may photoemit within $p_{T,max}=\sqrt{m_0\Delta E}$ (dashed lines) for the transverse (010)/(001) and (110)/(1$\overline{1}$0) crystal directions. (b) Transverse momentum distribution of the photoemitted electrons in the (010)/(001) and (110)/(1$\overline{1}$0) directions. (c) Incident photon energy dependence of the rms transverse momentum $\Delta p_T$ for electron temperatures $T_e \to 0$ (data points with solid line fit); $\Delta p_T=A\sqrt{m_0\Delta E}$, $T_e=$300K and the melting points of each metal (dot-dashed lines), together with the expected form of $\Delta p_T=\sqrt{(\hbar\omega-\phi)/3}$ (dashed line).}
\label{fig:Cucontour_001_001.pdf}
\end{figure*}
\\

\begin{table}[htb] 
\setlength{\tabcolsep}{5pt}
\caption{DFT evaluated photoemission properties of Cu and Au.}
\centering 
\begin{tabular}{l l l l l l}  
\hline\hline 
Metal & face & $\Delta p_{T,DFT}$ ($\sqrt{m_0eV}$) & $\Delta p_{T0}$ ($\sqrt{m_0eV}$) & $A_{T=0K}$\\
\hline
Ag& (100) &0.202 & 0.235 & 0.860 \\
  & (110) &0.202 & 0.235 & 0.860 \\
Cu& (100) &0.170 & 0.191 & 0.445 \\
  & (110) &0.198 & 0.288 & 0.444 \\
Au& (100) &0.208 & 0.278 & 0.425 \\ 
  & (110) &0.168 & 0.361 & 0.245 \\
\hline
\end{tabular} 
\label{tab:noblept} 
\end{table}

The DFT-based photoemission model described in the previous papers~\cite{li2017pbte, li2016photoelectric, bcc_jap, :/content/aip/journal/apl/101/19/10.1063/1.4766350, PhysRevSTAB.18.073401} is used to evaluate $\Delta p_T$ for Ag, Cu, and Au. The results for photoemission from the (001) and (110) faces of Cu and Au are shown in Fig.~\ref{fig:Cucontour_001_001.pdf}. The spatially-averaged values of $\Delta p_T$ extracted from the excess energy contours are calculated using $\hbar\omega$=4.75 eV for Cu and $\hbar\omega$=5.77 eV for Au, and the results are listed in~\ref{tab:noblept}. The theoretical data is a good fit to $\Delta p_T=A\sqrt{m_0(\hbar\omega-\phi)}$, giving the $A$ values within 79\% of 0.577 for Ag all cases expect Au(110) where $A$ value within 42\% of 0.577. In the $\Delta p_T(\hbar\omega, T_e)$ plot (\ref{fig:Cucontour_001_001.pdf}(c)), the rms transverse momentum is evaluated under 0 K, 300 K and the melting point, where the melting temperature for Cu (Au) is 1363K (1336K)~\cite{cohen_melting_1966}. As would be expected, the increase in $\Delta p_T$ due to electron temperature increase is weak (strong) for Au (Cu) which has the large (small) excess photoemission energy; both metals have inward bent `electron-like' energy states. Owing to the high degree of symmetry of Fermi surface in the FCC Brillouin zone, a relatively isotropic transverse momentum distribution is dominant in the prevalent crystal orientations.\\

\section{Photoemssion from the Surface State} \label{sec:Hollow Cone Illumination}
\begin{figure}[ht!]
\caption{FCC (111) Surface Brillouin Zone}
\vspace{-0.00cm}
\hspace{-0.00cm}
\centering
\includegraphics[scale=0.8]{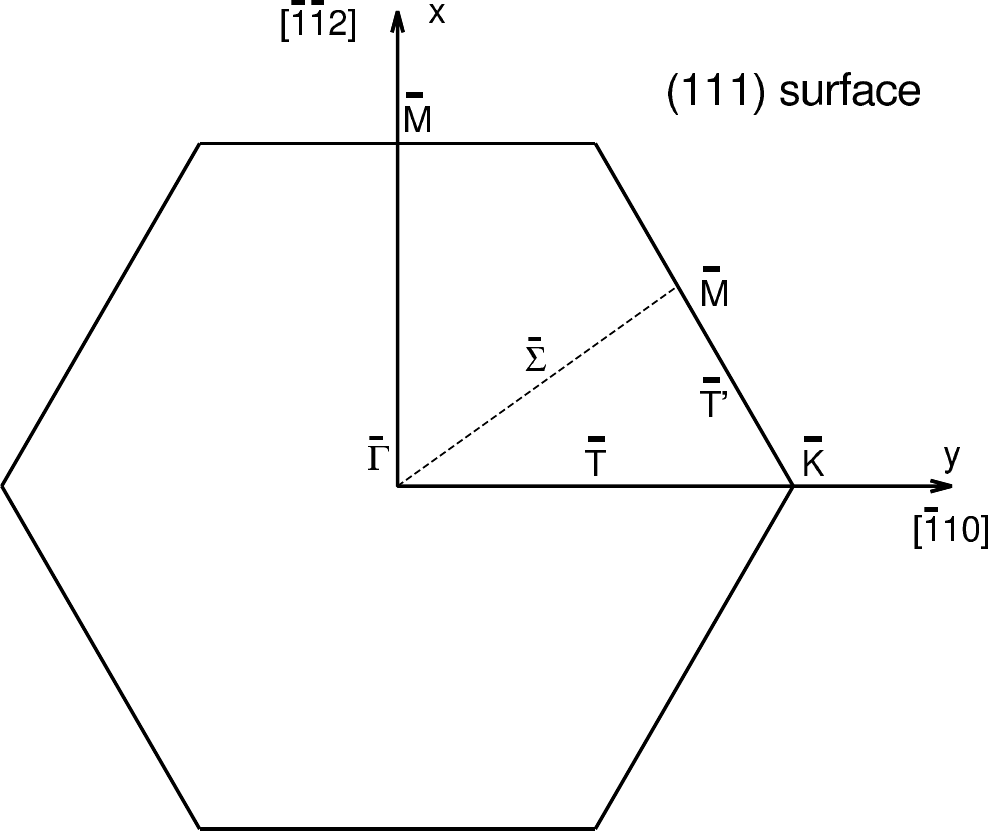}\\
\label{fig:sbz.pdf}
\end{figure}

Previous literatures have indicated that there are surface states in the Surface Brillouin Zone (SBZ) for Ag(111), Cu(111) and Au(111)~\cite{paniago_temperature_1995, burgi_noble_2000, kevan_high-resolution_1987, nicolay_spin-orbit_2001}. Due to the presence of a crystal surface, bulk-forbidden electronic single-particle states may arise leading to a band in the corresponding projected bulk band gap. These so-called surface states are highly localized perpendicular to the surface, and form a quasi two-dimensional electron band. Typically, surface states exist if their character is similar to a bulk state, but shifted in energy by the surface perturbation. Because the occupied part of the surface state is far away from the surface Brillouin-zone (SBZ) boundary, its dispersion can be regarded as identical for all in-plane directions. As shown in \ref{fig:sbz.pdf}, the SBZ is a two dimensional hexagonal structure for the (111) FCC crystal face ($k_z$ = 0), and it is centered at $\overline{\Gamma}$ point with $\overline{\Gamma}\overline{M}$ and $\overline{\Gamma}\overline{K}$ directions as $\vec{k_x}$ and $\vec{k_y}$, respectively; that is to say, the surface state disperses with momentum parallel to the surface $p_x$ and $p_y$ but not with perpendicular momentum $p_z$ (momentum is restricted with the in-plane vector $\vec{p_T}=\vec{p_x}+\vec{p_y}$). According to Ref. \cite{burgi2000noble}, \cite{PhysRevB.49.13897}, and \cite{nicolay_spin-orbit_2001}, it is the surface state which can play an essential role in noble metal (111) face photoemission because, for a clean (111) surface, the photoemission intensity reaches a speak at $E_F$ - $E_{ss,min}$, where $E_{ss,min}$ is the surface state band energy minimum~\cite{goldmann_empty_1985}. Surface states can therefore dominate emission from the (111) faces of noble metals.
\\

A successful method of distinguishing surface state and bulk photoemission features and determine the former's properties is to use a DFT-based surface band calculation within the slab model to extract the slab's band structure and inspect its energy dispersion properties, and thereby find the surface state band that is located between Fermi level and bulk bands. First, a (111)-face supercell comprising the multi-layer slab and suitable vacuum region needs to be constructed. Second, band structure calculation along the $\overline{\Gamma M}$ and $\overline{\Gamma M}$ directions is performed. Third, the bulk bands are projected onto the $k_x$-$k_y$ plane and plotted together with surface bands to determine the allowed surface states. As the surface energy dispersion relations are described in the surface Brillouin zone (See Fig.~\ref{fig:sbz.pdf}), in order to estimate where a surface state is, one can project the bulk bands into the two-dimensional Brillouin zone and then look for gaps that might accommodate such a state among the bulk states. The surface bands are the highest-energy partially occupied bands that fall below the Fermi level $E_F$ only for limited regions of the momentum-space. The electronic states of these surface bands are usually well localized in the vicinity of the surface. The electrons occupying such a region of allowed $k_x-k_y$ space require the least energy to overcome the summation of the work function and transverse energy required to escape the material as photoelectrons. When the laser photon energy is high enough to liberate these electrons but insufficient to reach the next energy band, an electron beam suitable for HCI can then be generated by emission from the single assessed surface state.\\

\begin{figure*}[!htb]
\setlength{\tabcolsep}{0pt}
\centering
\begin{tabular}{@{}cc@{}}
\includegraphics[scale=0.8]{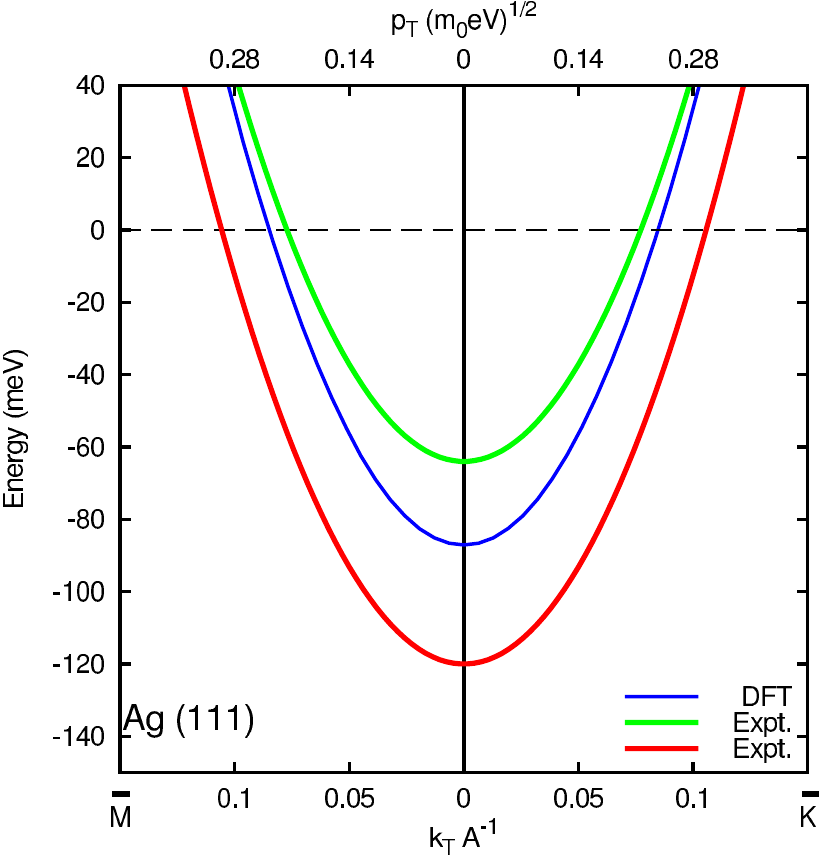}&
\includegraphics[scale=0.76]{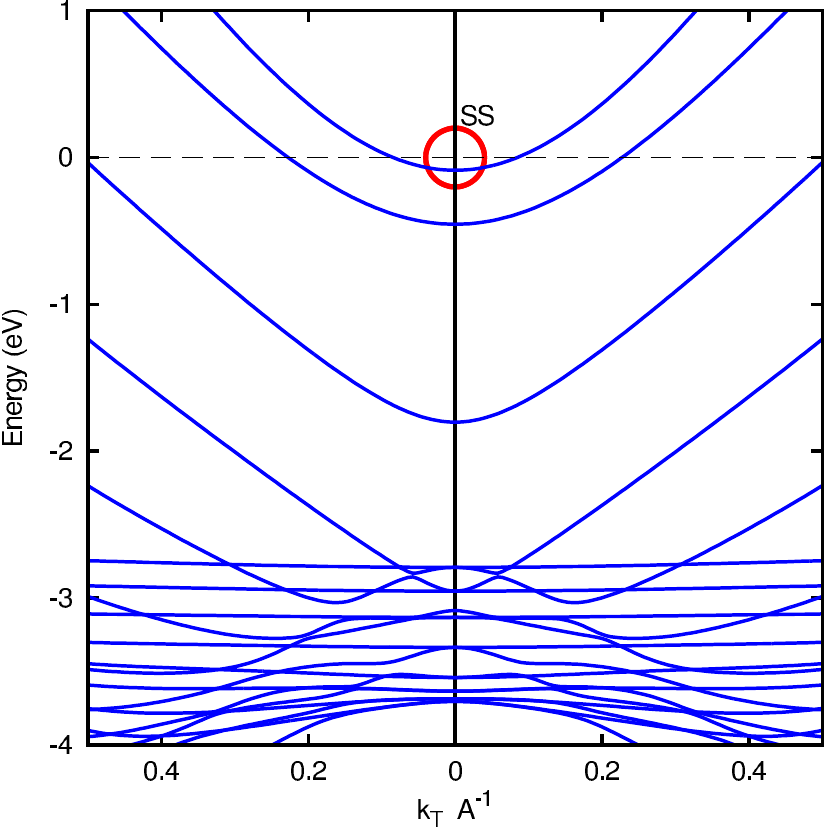}\\
(a) & (b)\\
\end{tabular}
\caption{Energy dispersion relation for the Ag(111) Surface state. Results of the band-structure calculation along the $\overline{\Gamma} \overline{M}$ direction for a 9-layer slab of Ag(111). The Fermi level has been adjusted to zero level. The left panel shows the comparison between the DFT calcuated surface band (blue solid line) with the experimental measured surface bands. The blue solid line is Nicolay's result (red solid line); the red solid line is Paniago's result (green solid line). The right panel shows the surface dispersion and the surface state is highlighted by red circle.}
\label{fig:Agsurface.pdf}
\end{figure*}
For Ag(111), there is an $s-p$ gap at $\overline{\Gamma}$, where a Shockley-type surface state was observed by R.Paniago, G.Nicolay et al~\cite{paniago_temperature_1995, nicolay_spin-orbit_2001, kevan_high-resolution_1987}. The $L$-gap $s-p$ surface state dispersion was obtained by measuring the energy of the surface state as a function of the external electron angle $\theta$. In contrast to the experimental investigations, we use a well-converged basis set and a $k$-mesh of 17$\times$17$\times$1 points in the two-dimensional irreducible Brillouin zone and modeled the surface by employing a periodic slab of 9 atomic layers separated by 15 $\AA$ of vacuum. The thin slab band calculation~\cite{fall_deriving_1999} is implemented by quantum-ESPRESSO using the local-density approximation~\cite{PhysRevB.41.7892} pseudopotential. Results of the band structure calculation along the $\overline{\Gamma  M}$ direction for the 9 layer slab of Ag(111) is shown in Fig.\ref{fig:Agsurface.pdf}(b). In order to be consistent with the experimental measurement unit, we plot the band dispersion in terms of E($\vec{k}$). In Fig.\ref{fig:Agsurface.pdf}(b), the blue solid lines give the slab's band dispersion by DFT at $T_e$ = 0, and the Fermi level has been adjusted to zero (dashed line), the allowed surface surface is highlighted by the red circle. In Fig.\ref{fig:Agsurface.pdf}(a), we compare my DFT calculation (solid green line) with the experimental Ag(111) surface state measurement of Nicolay~\cite{nicolay_spin-orbit_2001} (blue solid line) and by Paniago~\cite{paniago_temperature_1995} (red solid line). The DFT calculated surface state is between the two experimental surface states, which again demonstrates that there is a good agreement between experiment and theory.\\

The surface state, with its parabolic dispersion about the $\overline{\Gamma}$ point of the SBZ, is 62 meV below the Fermi level with effective mass $m^*$ = 0.40 $m_0$ in both the $\overline{\Gamma}$ $\to$ $\overline{\text{K}}$ and $\overline{\Gamma}$ $\to$ $\overline{\text{M}}$ directions, and has a Fermi wavevector value of $k_F$ = $\pm$0.08 $\AA^{-1}$~\cite{paniago_temperature_1995}. The DFT-based surface state dispersion shown in Fig.\ref{fig:Agsurface.pdf} gives the surface state band minimum $\sim$70 meV below $E_F$, and $m^*/m_0$ = 0.42. The DFT evaluated occupied part of the surface state ($k_F$ = 0.077 $\AA^{-1}$) is far away from the surface Brillouin-zone boundary ($k_{\overline{\Gamma} \overline{M}}$ = 0.577 $\AA^{-1}$), and its dispersion can be regarded as identical for all in-plane directions. Therefore, we restrict most of our following discussion to the dispersion along the $\overline{\Gamma}\overline{M}$ and $\overline{\Gamma}\overline{K}$ directions, denoted as the $\vec{p_x}$ and $\vec{p_y}$ directions with the in-plane vector $\vec{p_T}=\vec{p_x}+\vec{p_y}$. The minimum energy required to photoemit from the surface state, $E_{(111)}$, is the sum of FCC (111) face work function $\phi_{(111)}$ and the (111) Fermi level transverse momentum offset $p^2_F/2m_0$;
\begin{equation}\label{eq:m*m01}
E_{(111)}=\phi_{(111)}+p_F^2/2m_0,
\end{equation}
as in this case, $m^*$ $<$ $m_0$. Consequently, since $p_{F}$ = 0.245 $\sqrt{m_0eV}$, photoemission from the Ag(111) surface state is predicted to only occur when $\hbar\omega$ $\textgreater$ $\phi_{(111)}$ + 0.03 eV, where $\phi_{(111)}$ = 4.83 eV. This means that photoemission from the surface state of Ag(111) will first occur from states around the Fermi level as this requires the least energy. 
\begin{figure*}[ht!]
\centering
\begin{tabular}{@{}cc@{}}
\includegraphics[scale=0.51]{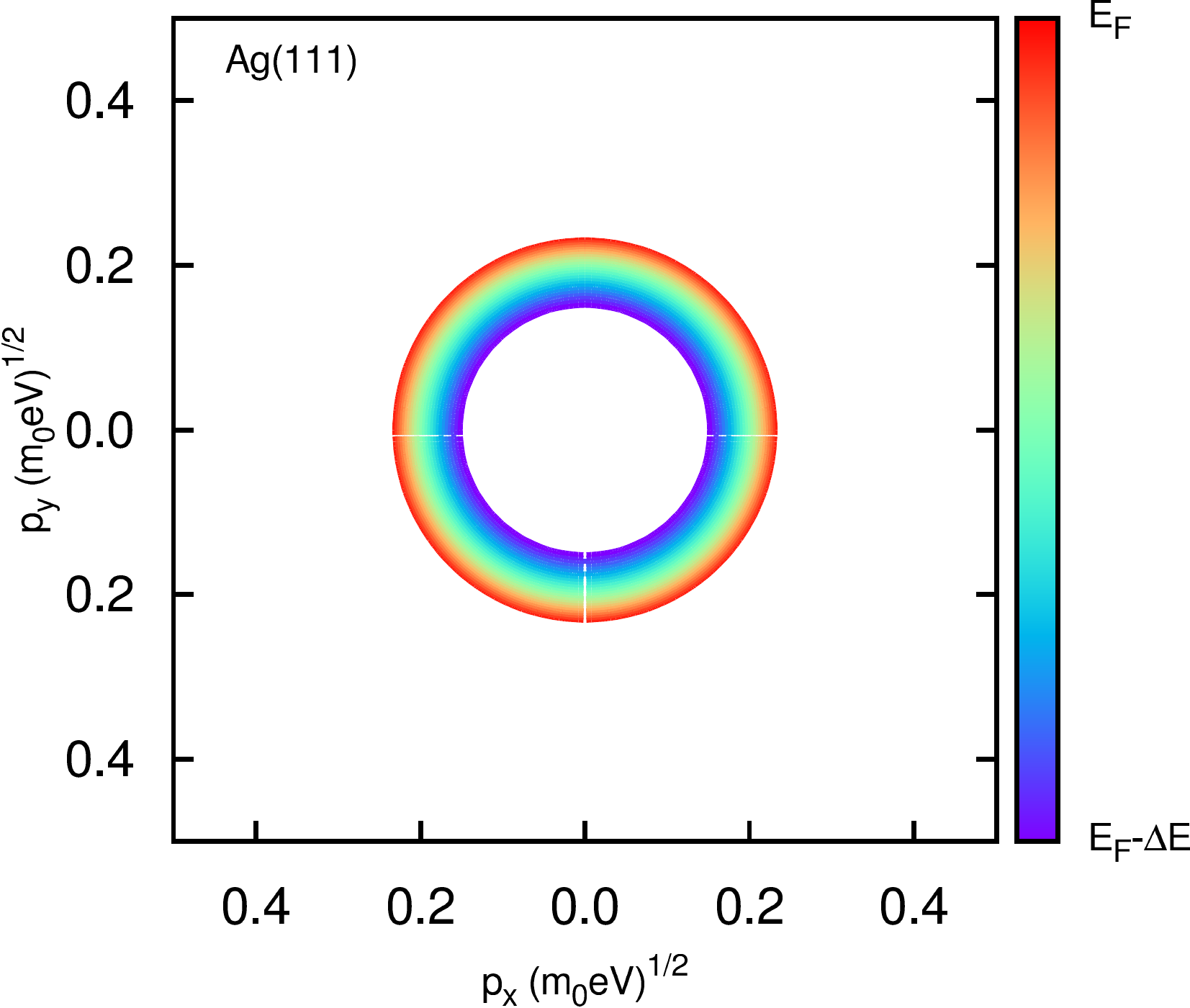}&
\includegraphics[scale=0.9]{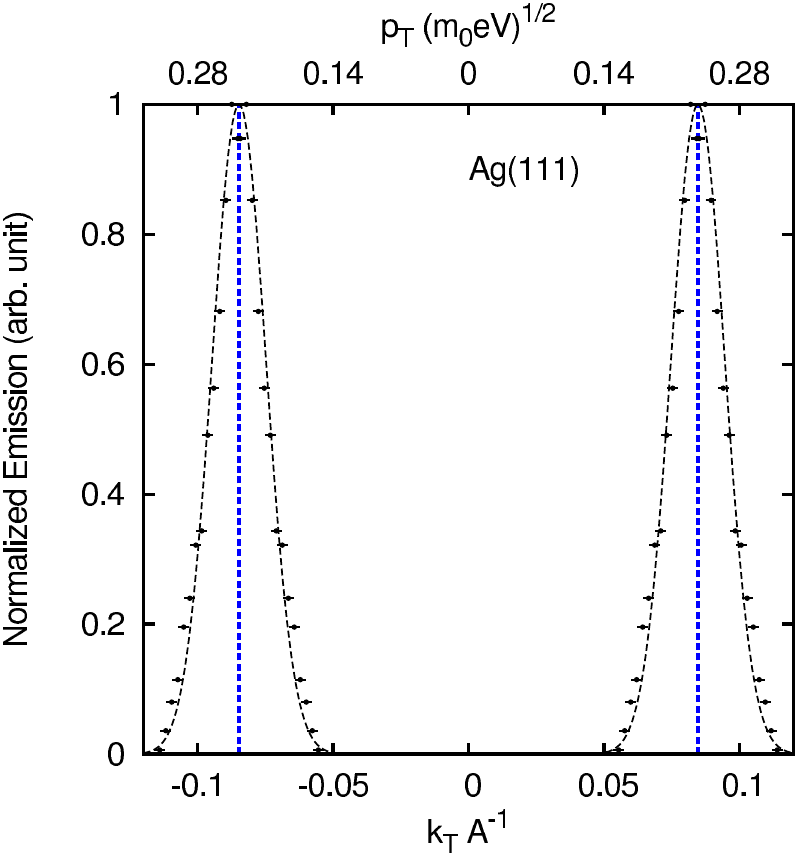}\\ 
(a) & (b) \\ 
\end{tabular}
\caption{(a) Ag(111) surface state constant energy contour with $\Delta E$ = 35 meV. The outmost red contour is the two dimensional surface state Fermi surface contour with radius 0.220$\sqrt{m_0eV}$. The innermost blue contour is the $E_F-\Delta E$ constant energy contour with radius 0.190$\sqrt{m_0eV}$. (b) Ag(111) surface state transverse momentum distribution with $\Delta E$ = 35 meV. Dashed line Gaussian fits are guides to eyes.}
\label{fig:Ag111band.pdf}
\end{figure*}
\\

From energy dispersion point of view, the surface state band structure is a single upward parabola; that is to say, the two dimensional Fermi surface map in the $p_x$ --- $p_y$ plane consists of only one single circle and at the maximum $p_T$ value --- $\hbar k_F$. For an excess energy of $\Delta E = 35$ meV, the resulting energy contour for the Ag(111) surface state has the `ring-like' shape as shown in Fig.~\ref{fig:Ag111band.pdf}(a); the two dimensional photoemitting energy contours reside inside the $L$-gap of the projected bulk band states in $p_T$-space.\\

The surface state photoemission energy-momentum relationship for an one-step photoemission process can be written as
\begin{equation}\label{eq:surface state boundary conditions}
\frac{(p^2_x+p^2_y)}{2m_0}+\phi_{(111)}+\frac{p^2_{z0}}{2m_0}=\hbar\omega+E(p_x, p_y),
\end{equation}
where $p_x$ and $p_y$ are the transverse momentum components ($p_T=\sqrt{p^2_x+p^2_y}$), $p_{z0}$ is the longitudinal momentum in vacuum and E($p_x$, $p_y$) is initial electron state energy. An intrinsic emittance calculation may thus be obtained from the transverse momentum distribution of photoemitting energy contours. After application of Eq.~\ref{eq:surface state boundary conditions}, the resulting transverse momentum distribution displayed in Fig.~\ref{fig:Ag111band.pdf}(b) shows the $L$-gap surface state photoemission. The $\Delta p_T$ of the `ring-like' emission peaks is 0.029$\sqrt{m_0eV}$ for $\Delta E$ = 35meV. It can be seen that the transverse momentum distribution is symmetrical with respect to $\overline{\Gamma}$ and has maxima close to $p_F$, which is consistent with experimental measurements.\\

As the emitted transverse momentum distribution is in the form of a ring with a radius $k_F$ = 0.077$\AA^{-1}$, its acceleration in an electron gun to momentum of $p_0$ will generate a hollow cone beam with a semi-angle $\theta_{HCI}$ $\approx$ $p_F/p_0$. That is, using the DFT-based thin-slab model, we show that the Ag(111) surface state photoemission with an excess energy of 35meV will produce a HCI beam. The useful range of excess energies for Ag(111) generation of HCI is 0 $\sim$ 62 meV, which will require accurately tunable UV laser; Ag(111) alloys such as Cu-Ag (111) could overcome this hurdle. It is also important to mention that when the surface state band effective mass is greater than the free electron mass, photoemission from the surface state will first occur from states around the $\overline{\Gamma}$ point as this requires the least energy. \\

\section{Summary and Discussion} \label{sec:summary and discussion}
This paper describes the emission properties of three noble metals (Ag, Cu, and Au) and the emission from the surface state. The Fermi surface, effective mass, and band structures for the (100), (110), and (111) faces are evaluated to confirm and supplement the current dataset regarding the electronic properties of the four FCC metals. The nonparametric model discussed can be generalized, implemented, and extended toward applications such as material design, data analysis, and multivariate modeling~\cite{wang2017methodology}. A surface state DFT evaluation has performed to show that the photocathode generated hollow cone illumination (HCI) can be realized through Ag(111) single crystal with 0.03eV excess energy. In case of $m_0$ $>$ $m_*$, the minimum energy required to photoemit from the surface state, $E_{(111)}$, is the sum of FCC (111) face work function $\phi_{(111)}$ and the (111) Fermi level transverse momentum offset $p^2_F/2m_0$ (\ref{eq:m*m01}). The $\Delta p_T$ of the `ring-like' emission from Ag(111) surface state peaks is 0.029$\sqrt{m_0eV}$.\\

\begin{acknowledgments}
The authors are indebted to Juan Carlos Campuzano, Christopher Grein, Randall Meyer, Alan Nicholls, and Serdar \"{O}\u{g}\"ut for their valuable discussions. This work was partially supported by the Department of Energy (Contract No. DE-FG52-09NA29451), Futurewei Techonologies, Inc., and Uptake Technologies, Inc.
\end{acknowledgments}
\bibliography{PhD.bib}
\end{document}